\documentclass[12pt]{JHEP3}
\usepackage{epsfig}

\usepackage{amsfonts}
\usepackage{amssymb}

\newcommand{\be}{\begin{equation}}
\newcommand{\ee}{\end{equation}}
\newcommand{\bea}{\begin{eqnarray}}
\newcommand{\eea}{\end{eqnarray}}
\newcommand{\ba}{\begin{array}}

\newcommand{\ea}{\end{array}}

\def\bbox{{\,
\lower0.9pt\vbox{\hrule \hbox{\vrule height 0.2 cm
\hskip 0.2 cm \vrule height 0.2 cm}\hrule}\,}}
\newcommand{\dsl}{\pa \kern-0.5em /}
\renewcommand{\t}{\theta}

\newcommand{\nn}{\nonumber\\}

\def\o{\omega}
\def\th{\theta}

\def\ds{\raise.15ex\hbox{/}\kern-.57em\partial}
\def\Ds{\,\raise.15ex\hbox{/}\mkern-13.5mu D}
\def\d{\delta}
\def\l{\lambda}
\def\o{\omega}
\def\t{\tau}

\def\o{\omega}

\def\pa{\partial}
\def\bzzr{|B_{ZZ}\rangle}
\def\blr{|B_{\l}\rangle}
\def\bzzl{\langle B_{ZZ}|}

\def\b{\beta}
\def\Th{\Theta}

\newcommand{\beq}{\begin{equation}}
\newcommand{\eeq}{\end{equation}} \newcommand{\beqn}{\begin{eqnarray}}
\newcommand{\eeqn}{\end{eqnarray}}

\preprint{
ITFA-2003-56\\
KIAS-P03089\\
hep-th/0312135\\
}

\title{\Large\bf String Interactions in $c=1$ Matrix Model}

\author{Jan de Boer, Annamaria Sinkovics, Erik Verlinde
and Jung-Tay Yee
\footnote{Also at the Korea Institute for
Advanced Study, Seoul, Korea}\\
email: jdeboer@science.uva.nl, sinkovic@science.uva.nl,
erikv@science.uva.nl, jungtay@science.uva.nl \\
\vskip 0.2mm
Institute for Theoretical Physics, University of Amsterdam \\
Valckenierstraat 65, 1018 XE Amsterdam, The Netherlands}

\abstract{We study string interactions in the fermionic formulation
of the $c=1$ matrix model. We give a precise nonperturbative
description of the rolling tachyon state in the matrix model, and discuss
S-matrix elements of the $c=1$ string. As a first step to study string
interactions, we compute the interaction of two decaying
D0-branes in terms of free fermions.
This computation is compared with the string theory cylinder diagram
using the rolling tachyon ZZ boundary states.}

\begin{document}

\section{Introduction}

It has been known for more than ten years that the non-critical
$c=1$ string can be described by the double scaling limit of a
quantum mechanical matrix model, for reviews see e.g.
\cite{Ginsparg:is,klebanov,polchinski}. In this duality the matrix
model describes discretized world-sheets, which become continuum
world-sheets in the double scaling limit. Recently, a new
perspective on this duality has emerged which is in spirit closer
the AdS/CFT duality and open/closed string duality. The crucial
new insight obtained in \cite{mcgrverl1,mcgrverl2} and in
\cite{klebanovmaldacenaseiberg} is that the matrix degrees of
freedom of the quantum mechanical matrix model are to be
identified with the open string tachyon degrees of freedom living
on a system of unstable D0-branes in the $c=1$ string theory. The
full $c=1$ string theory is recovered by taking a double scaling
limit of the matrix model. For more discussion see e.g. the recent
papers \cite{many1,cola,gutperlekraus, newhat, takayanagi,
dewolfe}.

The $c=1$ matrix quantum mechanics can be reformulated in terms
of free fermions moving in an external potential.
After taking the double scaling limit the potential becomes an
upside-down quadratic potential, the system of free fermions becomes
a non-relativistic $1+1$ dimensional field theory, and the ground state
is a Fermi sea filled with
an infinite number of fermions. The distance $\mu$ of the Fermi sea to the top
of the potential can be identified with the inverse string coupling
${1 \over g_s}$ \cite{klebanov}.

It is an interesting question to ask how the string interactions
and string diagrams are encoded in the free fermionic theory.
In principle, one expects that different orders in string perturbation
theory can be extracted from the matrix model by studying its large
$\mu$ expansion.
The tree-level scattering of closed string tachyons in the two dimensional
string theory were computed in the matrix model (where one can compute
them directly or by employing the free fermionic formulation)
in \cite{klebanov,klebgross}. These scattering amplitudes
were found to be in agreement with closed string computations
when on the matrix model side external ``legpole factors'' are attached.
These factors have poles at integer values of the external
momenta, corresponding
to discrete states in the string theory. After analytic continuation of the string amplitudes to Minkowski
signature, these factors turn into a phase, which do not affect the physical
scattering amplitudes.

However, the discrete states and the legpole factors play an
important role in understanding the full structure of the $c=1$
string theory. As shown in \cite{polchinski} gravitational effects
are encoded in the legpole factors.

In the $c=1$ matrix model the full perturbative S-matrix was
also computed  \cite{polchinski, mooreplramg}. Nonperturbatively
this matrix model is unstable, since only one side of the potential
is filled, and there are no closed-string states corresponding to
the asymptotic region on the other side of the potential.
A non-perturbatively unitary extension of the $c=1$ matrix model
is obtained by filling up both sides of the potential, in which
case it does not describe the non-critical $c=1$ string but the
two dimensional 0A/0B string theory \cite{newhat}.
In this unitary theory an exact
S-matrix can be computed \cite{dewolfe, mooreplramg}.

The partition function of the two dimensional string theory and
its Kontsevich representation was worked out in \cite{DMP}.

Although many results have already been computed
on the $c=1$ matrix model side,
in the present context we have a better understanding of the direct relation
of the matrix model to the D0-branes of the two dimensional string theory.
In particular, processes involving D-branes on the string side
can now be studied with matrix model methods.

In \cite{mcgrverl1, mcgrverl2, klebanovmaldacenaseiberg} the decay of
a single D-brane was analyzed in detail. In this picture, the decay of
the D0-brane can be described by a single eigenvalue rolling down
the potential. In the fermionic language one can describe this process
by a single fermion moving from the top of the potential to asymptotic
infinity, where it becomes relativistic.
Bosonizing the relativistic fermion leads to the identification
of the D-brane as a coherent state of tachyons in the asymptotic region.
The closed string interactions are encoded in the interactions of the
bosonized tachyon field.
A general open-closed duality based on this description is suggested
in \cite{senopenclosed}.

In the string theory side, the decay of the unstable D-brane can be
studied using a description involving
boundary conformal theory. The relevant boundary states have two non-trivial
pieces, one is the Liouville part which has been exactly determined in
\cite{FZZ,ZZ}, the other is the time-dependent piece which involves
a timelike scalar and which was first studied in \cite{senroll}.
A precise determination of the time-dependent part of the boundary
state is actually quite subtle and we will discuss it in more detail
later in this paper.

As a first step in understanding in more detail the relation between the
string worldsheet description and the dual
matrix model, one would like to see how the string interactions
are reproduced in the matrix model. In particular, since
the matrix model has a description in terms of a simple system of
free fermions in a quadratic potential, one would like to understand
if and how string interactions between D-branes can be computed
from the free fermionic theory.

So far, in various papers \cite{ mcgrverl1,
mcgrverl2, klebanovmaldacenaseiberg, liulambertmaldacena}
the disk one point function describing the
decay of a single D-brane was computed and found to agree
with the bosonized fermionic description.
A further step is made in \cite{gutperlekraus}, where higher
$g_s$ corrections to the outgoing tachyon state were discussed using
the S-matrix of the of the matrix model.

In the present paper we further analyze the question of string
interactions in the presence of D-branes from the free fermionic
point of view.
The main point is to understand and compute
string diagrams (the disk and the cylinder), including possible
higher order and nonperturbative corrections,
in terms of the free fermionic formulation of the matrix model.

In section \ref{MatrixI} and \ref{MatrixII} we develop the systematic
formalism to compute string diagrams from the free fermionic point of view.
The disk diagram, which has been computed before serves mainly as an example
to develop our formalism. As described in section \ref{MatrixII}
the rolling tachyon can be viewed as a combination of in and out states.
Here we use the S-matrix formalism to compute the disk amplitude and
the string interactions between D0-branes.
We conclude that
the matrix model results are in agreement with the boundary state computations,
if in the latter we take into account the exchange of closed string
tachyons only.
In particular, we do not see discrete modes contributing in the matrix
model computation.

To compare the matrix model formulation of the interaction
 with the boundary state formulation,
in section \ref{cylinder},
we compute the tree level interaction
between two decaying D0-branes.
The Liouville part of the cylinder diagram
is already computed in \cite{ZZ, martinec}. Tensoring with a
Dirichlet state in the time direction describes for example the
cylinder diagram between two D-instantons \cite{newhat, takayanagi}.
Here we consider the tree-level interaction between two decaying
D0-branes.
We will take as our boundary state the one that arises by an analytic
continuation of the zero mode sector of the Euclidean boundary state,
and briefly discuss possible alternatives to this.

The comparison of the boundary state computation
 and matrix model results is further
discussed in section \ref{discussion}.

\section{The interaction of D-branes : A bosonic picture} \label{MatrixI}

In this section, we briefly review some background material pertaining to the
$c=1$ matrix model and the asymptotic bosonization
picture of free fermions. Then we explain how to reproduce the brane
``interaction'' from this picture. The Hamiltonian of free
fermions in the inverted harmonic oscillators with Fermi
energy $\mu$ is given by
 \bea
 \label{Hamiltonian}
  H = {1 \over 2} \int dx \left\{
  \partial_x \psi^\dagger \partial_x \psi - x^2 \psi^\dagger \psi
  + 2 \mu \psi^\dagger \psi \right\}.
 \eea
 We can introduce new left moving and right moving chiral
fermions
 \bea
 \label{chiralfermion}
  \psi(x,t) = {1 \over \sqrt{2 v(x)}} e^{+i \mu \t-i \int^x dy v(y)}
  \psi_L(x,t) + {1\over \sqrt{2 v(x)}} e^{+i \mu \t+i \int^x dy v(y)}
  \psi_R(x,t),
 \eea
where
 \bea
  v(x) = \sqrt{x^2 - 2 \mu}.
 \eea
The factors appearing in front of $\Psi_{L}$ and $\Psi_{R}$ are the WKB
wavefunctions of the inverted harmonic oscillator. From now on, we
will focus on the asymptotic region of large negative $x$, or the $q
\equiv -\ln(-x) \rightarrow -\infty$ region. The reason for this will
soon become clear.  At this asymptotic
infinity, the Hamiltonian is reduced to the Hamiltonian of relativistic
fermions and we have
 \bea
 \label{chiralhamiltonian}
  H= {1 \over 2} \int d q \left[ i \psi_R^\dagger \partial_q \psi_R
   - i \psi_L^\dagger \partial_q \psi_L + {\cal O}\left({1\over \mu}
   \right) \right].
 \eea
Here we suppressed $1/\mu$ corrections. This can be justified
for the tree level computation in string theory. Possible systematic
correction to this truncation will be discussed in section \ref{MatrixII}.
The chiral fermions can be bosonized
 as
 \bea
 \label{bosonization}
  \psi_R(q,t) &=& {1 \over \sqrt{2 \pi}} : \exp\left[{i \over
     \sqrt{2}} \int^q
  (\Pi- \partial_q T) d q' \right]: \nn
  \psi_L(q,t) &=& {1 \over \sqrt{2 \pi}} : \exp\left[{i \over
     \sqrt{2}} \int^q
  (\Pi+ \partial_q T) d q' \right]:.
 \eea
This scalar boson is, roughly speaking, a closed string tachyon modulo
possible discrete modes.
The Hamiltonian of this scalar boson can be found easily by inserting
(\ref{bosonization}) into (\ref{chiralfermion}) and evaluating
(\ref{Hamiltonian}). It has interactions
of order $O(T^3)$ which are again suppressed by inverse powers of $\mu$,
 \bea
  \label{bosonicHamiltonian}
  H={1 \over 2}\int dq \left( \Pi^2+ (\partial_q T)^2 + O(T^3)\right).
 \eea
The field $T(q,t)$ satisfies the canonical equal time commutation
relations
 \beq
  \left[ T(q,t), \Pi(q',t) \right] = i \delta(q- q').
 \eeq
Near asymptotic infinity, the bosonic field $T$
can be expressed in terms of simple plane
waves,
 \bea \label{aux11}
  T(q,t) = \int_{-\infty}^{+\infty} { d k \over \sqrt{2 k^2}}
  \left[ T_k e^{-i |k| t+ i k q} + T_k ^\dagger e^{i |k|t - i k q}
  \right].
 \eea
The commutation relations of the creation and annihilation
operators of this bosonic field are the usual ones,
 \bea
  [T_k, T_{k'}^\dagger] = |k| \delta(k-k') \label{comm}.
 \eea
Including the normal ordering, (\ref{bosonization}) can now be
written as
 \bea
 \label{bosonization2}
  \psi_R(q,t) &=& {1 \over \sqrt{2 \pi}} e^{- i
     \int_0^{+\infty}{dk\over |k|} {\bar T}_k^\dagger
     e^{+ik(t-q)}}e^{- i
     \int_0^{+\infty}{dk\over |k|} {\bar T}_k
     e^{-ik(t-q)}} \nn
  \psi_L(q,t) &=& {1 \over \sqrt{2 \pi}} e^{+ i
     \int_{-\infty}^0{dk\over |k|} T_k^\dagger
     e^{-ik(t+q)}}e^{+ i
     \int_{-\infty}^0{dk\over |k|} T_k
     e^{+ik(t+q)}}.
 \eea
We distinguish $T_k$ and ${\bar T}_k$, where the former acts naturally on the
$out$ vacuum and the latter naturally on the $in$ vacuum. The relation between
$T_k$ and ${\bar T}_k$ can be found by comparing and equating the
values of conserved quantities at very early and very late times. The
relevant conserved quantities are related by an analytic continuation
to the generators of the ground ring (which also correspond to conserved
quantities). The result, as taken from
\cite{polchinski}, reads
 \bea
 \label{mucorrection}
  {\bar T}_k^{\dagger} -{\bar T}_k &=& 2^{ik}
  \sum_{n=1}^\infty {1 \over n!} \left({i \over
  \mu}\right)^{n-1} {\Gamma(1-i k)\over \Gamma(2-n-i
  k)}\int_{-\infty}^0 dk_1\cdots dk_n \nn
  &&
  (T_{k_1}^\dagger- T_{k_1})
  \cdots
  (T_{k_n}^\dagger- T_{k_n}) \delta(\pm|k_1|\cdots \pm|k_n|-k).
 \eea
The leading one gives the tree level interaction, and since this
expression includes all $1/\mu$ corrections, we can include higher
order corrections \cite{gutperlekraus}.

States that describe the excitation of a single fermion out of the
Fermi sea can be expressed as
$\psi_R^\dagger(q,t)| \mu\rangle$ or
$\psi_L^\dagger(q,t)|\mu\rangle$, where $|\mu\rangle$ denotes the Fermi
sea. If we insert the classical trajectories of $x=-\hat\lambda
\cosh t$ or $q= -\log{\hat\lambda \over 2} - \log \cosh t$ in the
single fermion state, we get, for instance,
 \bea
   \psi_L^\dagger(q,t)|\mu\rangle = {1 \over \sqrt{2\pi}} e^{- i
   \int_0^\infty {dk\over |k|} T_k^\dagger e^{-ik\log
   \hat\lambda}} |\mu\rangle
 \eea
at asymptotic future infinity $t \rightarrow +\infty$. This single
fermion state can be interpreted as outgoing coherent tachyon states emitted
from the decaying D-brane \cite{klebanovmaldacenaseiberg}.

The question is how we can compute the ``interaction'' from the free
fermion theory. The strategy is as follows.  Although the theory is
free, there are non-vanishing multipoint functions of fermions,
 and we will interpret the two point function of fermions
$\langle \mu |\psi^\dagger (q,t) \psi(q,t)|\mu\rangle$ as an ``interaction
amplitude'' of D-branes.
We should note that since each fermion in the two
point function is interpreted as a decaying D-brane,
to get a sensible S-matrix element of two interacting D-branes
the $(q,t)$ coordinates should be sent to asymptotic infinity as
one usually does
when constructing an S-matrix.
Of course, this interpretation looks a bit different
from standard quantum field theory at this moment, but  in
the next section we will see that
once it is assembled together that the whole picture matches perfectly
well with standard quantum field theory.
At this asymptotic region, we can think about the ``interaction''
of two Fermi particles as
 \bea
  \langle \mu| \psi_L (q,t) \psi_L^\dagger (q',t') |\mu\rangle
     &=& \langle \mu| {1 \over \sqrt{2 \pi}} e^{-  i
     \int_{-\infty}^0{dk\over |k|} T_k^\dagger
     e^{-ik(t+q)}}e^{-  i
     \int_{-\infty}^0{dk\over |k|} T_k
     e^{+ik(t'+q')}}\nn
   &&\,\,\, \times {1 \over \sqrt{2 \pi}} e^{+ i
     \int_{-\infty}^0{dk\over |k|} T_k^\dagger
     e^{-ik(t+q)}}e^{+ i
     \int_{-\infty}^0{dk\over |k|} T_k
     e^{+ik(t'+q')}}  |\mu\rangle \nn
  &=& {1 \over 2 \pi} e^{+\int_0^{+\infty} {dk \over k}
  e^{-ik((t+q)-(t'+q'))}}
  ={1 \over 2 \pi}
  e^{+\int_0^\infty {dk\over k} e^{-ik
  (\log\hat\lambda-\log\hat\lambda')}
  } \nn
  \label{LL}
 \eea
where in the last step we inserted the classical trajectory $q=
-\log{\hat\lambda \over 2} - \log \cosh t$ in the correlation
function of two Fermi fields at asymptotic infinity. We have also
neglected the overlap of the wavefunctions of the fermions with
the Fermi sea in the computation. The form of
(\ref{LL}) suggests it is the ladder summation of cylinder diagrams. The single
cylinder can be rewritten as
 \bea
 \label{onecylinder}
  {\cal A}_{\rm cylinder} \sim \int dk d\omega
  e^{-ik
  \log\hat\lambda}{1 \over k^2-\omega^2}
  e^{+ik\log\hat\lambda'}.
 \eea
We see that (\ref{onecylinder}) is composed of three parts: a massless
scalar particle propagator and two one point functions. The scalar
particle is the closed string tachyon as we know from
(\ref{bosonization}). We will discuss further the match in the
section \ref{cylinder}, but already at this level, it is apparent
that even though (\ref{LL}) explains the closed string tachyon
exchange perfectly, there is no signal of discrete modes.

\section{The interaction of D-branes : A fermionic picture} \label{MatrixII}

We will now try to give a more precise definition of the rolling
tachyon state in the $c=1$ matrix model. Since the rolling tachyon
has a very specific behavior at early and late times as
we have considered in the previous section, we propose
that in the Minkowski picture it is related to a combination of
an in state and an out state. Bulk correlators in the $c=1$ matrix
model are evaluated between this in and out state. Both the in and
out state have a free parameter, which defines the behavior at
very early times $\sim a \exp(-X^0)$ and the behavior at very late
times $\sim b \exp (X^0)$. We will therefore not view the
parameters $a,b$ as defining a single state in the matrix model,
but as defining a pair of states consisting of
an in state and an out state. The in state
represents an incoming coherent wave of tachyons, the out state
and outgoing coherent state of tachyons, which agrees with the
intuitive picture of the rolling tachyon boundary state.

Of course, what we say here is basically the statement that a
classical fermion trajectory is nothing but a D-brane, but so far
in the literature the rolling tachyon has mainly been considered
as single state in the matrix model, rather than as a combination
of an in and an out state.
However, if we want to compute S-matrix
elements of the $c=1$ string, we need to specify both the initial
and final conditions of the D-brane with the rolling tachyon, and
we will see below that this is indeed the right picture.

The map between closed string tachyon correlation functions and
$c=1$ quantities has been discussed in e.g. \cite{klebanov}. It involves
the asymptotic bosonization described in the previous section,
in particular the modes in (\ref{comm}) are the modes of the closed
string tachyon.
The same chiral fermions that appear asymptotically will also be
used to construct the in and out states for the rolling tachyons,
but whereas closed string tachyons correspond to fermion
bilinears,  for the rolling tachyon we will use a single fermion
operator, as discussed in \cite{gutperlekraus}. The relation
between the exact $c=1$ fermion and the asymptotic chiral fermions
involves WKB factors \cite{klebanov} as in (\ref{chiralfermion})
that need to be removed in
order to obtain a finite result. This removal of WKB factors is
also exactly what is needed to obtain
finite tachyon correlators \cite{mooreplramg}.

The free fermion theory in the inverted harmonic oscillator has
standard even and odd parity real delta-function normalized
wavefunctions $\psi^{\pm}(\omega,x)$, as described in
\cite{mooreplramg, moore}. We want to find the wavefunctions
that have the appropriate plane wave behavior as $x\rightarrow
-\infty$. The relevant wavefunctions
are denoted $\chi^{\pm}_L(\omega,x)$; we could also have
considered the other regime $x\rightarrow +\infty$ with corresponding
wave functions $\chi_R(\omega,x)$ but will not do
that here. This will be relevant once we are interested in the
$\sinh$ boundary state (in the matrix model in which both sides of the
potential are filled), for example. The $\chi^{\pm}_L$ are linear
combinations of the standard wavefunctions, to be precise,
\be \label{eq1}
\left(
\begin{array}{c}
\chi_L^+(\omega,x) \\ \chi_L^-(\omega,x) \end{array}
 \right)
=
\left(\begin{array}{cc} s^-(\omega) & s^+(\omega) \\
s^+(\omega) & s^-(\omega) \end{array} \right) \left(
\begin{array}{c}
\psi^+(\omega,x) \\ \psi^-(\omega,x) \end{array}
 \right)
 \ee
with
\be s^{\pm}(\omega) =\sqrt{k(\omega)} \pm \frac{i}{\sqrt{k(\omega)}}
\ee
where
\be k(\omega) = \sqrt{1+e^{2\pi \omega}} -e^{\pi\omega},\qquad
k(\omega)^{-1} = \sqrt{1+e^{2\pi \omega}} +e^{\pi\omega} .
\ee
{}From (\ref{eq1}) we readily obtain
\be
\left(
\begin{array}{c}
\psi^+(\omega,x) \\ \psi^-(\omega,x) \end{array}
 \right) = \frac{i}{4} \left(\begin{array}{cc} s^-(\omega) & -s^+(\omega) \\
-s^+(\omega) & s^-(\omega) \end{array} \right) \left(
\begin{array}{c}
\chi_L^+(\omega,x) \\ \chi_L^-(\omega,x) \end{array}
 \right)
 \ee
 so that in terms of creation an annihilation operators (with
 standard commutation relations)
 we have
 \be
\psi(x,t) = \int d\omega \, e^{i\omega t} \left[ \frac{i}{4}
a_{\omega}^+ (s^- \chi_L^+ - s^+ \chi_L^-) +\frac{i}{4}
a_{\omega}^- (-s^+ \chi_L^+ + s^- \chi^-_L)\right] .
\ee
The asymptotics of $\chi^{\pm}_L$ as $x\rightarrow-\infty$ are
\be
\chi_L^{\pm}(\omega,x) \sim \frac{\mp 2i}{\left( 2 \pi |x|
\sqrt{1+e^{2\pi \omega}} \right)^{1/2}} e^{\pm i F(|x|,\omega)}
\ee
with
\be
F(x,\omega) = \frac{1}{4} |x|^2 - \omega \log|x| + \Phi(\omega)
\ee
with
\be
\Phi(\omega) = \frac{\pi}{4} +\frac{i}{4} \log \left[
\frac{\Gamma(\frac{1}{2} - i \omega)}{\Gamma(\frac{1}{2} + i
\omega)}\right] .
\ee
We are interested, for the in-state, in the operator
\be \label{instate}
\lim_{x \rightarrow -\infty} \psi^{\dagger}(t=-\log|x| +
\log(\hat{\lambda}_{\rm in}),x )
\ee
acting on the ground state $|\mu\rangle$. Similarly, for the out
state, we are interested in the operator
\be \label{outstate}
\lim_{x\rightarrow -\infty} \psi(t=\log|x| - \log(
\hat{\lambda}_{\rm out}), x )
\ee
acting on the vacuum state $\langle \mu |$. To obtain a finite resulting
expression, we need to restrict to $\chi_L^+$ (the appropriate
left/right moving part), and to remove the WKB piece
\be
\frac{1}{\sqrt{|x|}} \exp(\frac{i}{4} |x|^2) .
\ee
The resulting operators are most easily written in terms of the
operators
\bea \label{defop}
\Psi_{{\rm in},{\omega}}^{\dagger}
 & = & \frac{e^{i\Phi(\omega)-i\omega\log\sqrt{\mu}}}{2
(1+e^{2\pi\omega})^{1/4}} (s^-_{\omega} (a^+_{\omega})^{\dagger} -
s^+_{\omega}
 (a^-_{\omega})^{\dagger} )\nonumber \\
\Psi_{{\rm out},{\omega}} & = & \frac{e^{i\Phi(\omega)-i\omega\log\sqrt{\mu}}}{2
(1+e^{2\pi\omega})^{1/4}} (s^-_{\omega} a^+_{\omega} -
s^+_{\omega}  a^-_{\omega} )
\eea
which will play a crucial role in what follows. They describe the
modes of the asymptotic chiral fermions, but in terms of the {\it
exact} annihilation and creation operators of the $c=1$ theory.
The factors of $\log\sqrt{\mu}$ are put in by hand here in order to facilitate the
comparison with world-sheet results later.
The anti-commutation of the fermions  are
\bea
\{ \Psi_{{\rm out},{\omega}} , \Psi_{{\rm out},{\omega
'}}^{\dagger} \} & = & \delta(\omega-\omega') \nonumber \\
\{ \Psi_{{\rm in},{\omega}} , \Psi_{{\rm in},{\omega
'}}^{\dagger} \} & = & \delta(\omega-\omega') \nonumber \\
\{ \Psi_{{\rm in},{\omega}}^{\dagger} , \Psi_{{\rm out},{\omega '}} \} & = &
- \frac{e^{\pi\omega} e^{2i\Phi
(\omega)-i\omega\log\mu}
}{\sqrt{1+e^{2\pi\omega}}}
 \delta(\omega-\omega') \nonumber \\
\{ \Psi_{{\rm in},{\omega}} , \Psi_{{\rm out},{\omega
'}}^{\dagger} \} & = & - \frac{e^{\pi\omega} e^{-2i\Phi(\omega)+i\omega\log\mu}
}{\sqrt{1+e^{2\pi\omega}}} \delta(\omega-\omega').
\eea
The factor that appears in the the commutation relations between
the in and out modes is the reflection coefficient which is the
S-matrix of the free fermion theory. Most of the non-trivial
physics of the $c=1$ theory is contained in this reflection
coefficient. It can be rewritten as
\bea
R_{\omega} & = & - \frac{e^{\pi\omega} e^{2i\Phi(\omega)-i\omega \log\mu}
}{\sqrt{1+e^{2\pi\omega}}} \nonumber \\
&  = &  - \frac{i e^{\pi\omega-i\omega\log\mu}
}{\sqrt{1+e^{2\pi\omega}}} \sqrt{ \frac{\Gamma(\frac{1}{2} + i
\omega)}{\Gamma(\frac{1}{2} - i \omega)} } \nonumber \\
&  =  &
-\frac{i}{\sqrt{2\pi}} e^{\pi\omega/2-i\omega\log\mu} \Gamma(1/2+i\omega)
\eea
in view of the relation
\be
\Gamma(1/2 + i \omega) \Gamma(1/2 - i\omega) = \frac{\pi}{\cosh
\pi\omega} .
\ee
The factor $R_{\omega}$ decays exponentially at large negative
$\omega$, which guarantees that the results we obtain will be
finite. This exponential decay indicates that only
non-perturbative effects will access the region where the energy
is negative.

The reflection coefficient has nonperturbative pieces and a perturbative
expansion in $1/\mu$. The perturbative expansion is of the form
\be
R_{\mu+\omega} =- i e^{-i\mu} \left[ 1 + \frac{i}{2\mu}(\omega^2 + \frac{1}{12}) +
{\cal O}(\mu^{-2}) \right]
\ee
with an irrelevant prefactor $-i e^{-i\mu}$.

Based on our previous discussion, we now propose that the operator
that creates a D-brane with a rolling tachyon that as
$x\rightarrow -\infty$ behaves as
\be T \sim \hat{\lambda}_{\rm in} e^{-t}
\ee
is
\be
|{\rm in}\rangle  \equiv D^{\dagger}_{{\rm in},\hat{\lambda}_{\rm in}}
|\mu\rangle =
  \int d\omega\,
e^{-i\omega\log\hat{\lambda}_{\rm in}} \Psi_{{\rm in},\omega}^{\dagger}
 |\mu\rangle
\ee
and similarly that the out operator that annihilates a D-brane
with a rolling tachyon that as $x\rightarrow +\infty$ behaves as
\be T \sim \hat{\lambda}_{\rm out} e^{t}
\ee
is
\be
\langle {\rm out} |  \equiv \langle \mu | D_{{\rm out},\hat{\lambda}_{\rm out}}
=
  \langle \mu | \int d\omega\,
e^{-i\omega\log\hat{\lambda}_{\rm out}} \Psi_{{\rm out},\omega} .
\ee
Similarly, we can make in and out states that contain multiple
branes by simply acting with the corresponding brane creation and
annihilation operators.

Before continuing the discussion, we first test this proposal by
computing the energy associated to the rolling tachyon. In string
theory, this should be the one point function of the relevant
ground ring operator in the rolling tachyon boundary state. For
the present point of view the relevant calculation is easy to
write down, since the energy of $(a_{\omega}^{\pm})^{\dagger}$ is
$\omega$. The exact expression for the energy therefore is given
by computing the expectation value of $\omega$ between the in and
the out state, which after normalization leads to the expression
\be
E=\frac{\int_{-\infty}^{\mu} e^{-i \omega( \log\hat{\lambda}_{\rm
in} + \log\hat{\lambda}_{\rm out})} \omega
\frac{e^{\pi\omega}}{\sqrt{1+e^{2\pi\omega}}} \sqrt{
\frac{\Gamma(\frac{1}{2} + i \omega)}{\Gamma(\frac{1}{2} - i
\omega)} } }{ \int_{-\infty}^{\mu} e^{-i \omega( \log
\hat{\lambda}_{\rm in} + \log \hat{\lambda}_{\rm out} )}
\frac{e^{\pi\omega}}{\sqrt{1+e^{2\pi\omega}}} \sqrt{
\frac{\Gamma(\frac{1}{2} + i \omega)}{\Gamma(\frac{1}{2} - i
\omega)} }} .
\ee
In this expression we absorbed a factor of $\sqrt{\mu}$ in both
$\hat{\lambda}_{\rm in}$ and $\hat{\lambda}_{\rm out}$.

Let us try to work out this expression. In the region where
$\omega$ is close to $\mu$, the factor
$e^{\pi\omega}/\sqrt{1+e^{2\pi\omega}}$ can be dropped as it is
equal to $1$ plus nonperturbative corrections. We are left with
\be
E=\frac{\int_{-\infty}^{\mu} e^{-i \omega ( \log\hat{\lambda}_{\rm
in} + \log\hat{\lambda}_{\rm out})} \omega  \sqrt{
\frac{\Gamma(\frac{1}{2} + \omega)}{\Gamma(\frac{1}{2} - i
\omega)} } }{ \int_{-\infty}^{\mu} e^{-i \omega (
\log\hat{\lambda}_{\rm in} + \log\hat{\lambda}_{\rm out})} \sqrt{
\frac{\Gamma(\frac{1}{2} + i \omega)}{\Gamma(\frac{1}{2} - i
\omega)} }} .
\ee
If we assume the rolling eigenvalue has small overlap with the
Fermi sea, we can extend the integral to the range from $-\infty$
to $\infty$, and since it is rapidly fluctuating we can use the
method of stationary phase. The fluctuating determinant is
subleading in the genus expansion, so
we are interested in the leading stationary phase approximation. Using
Stirling's formula, the gamma functions behave as
\be
\sqrt{ \frac{\Gamma(\frac{1}{2} + i \omega)}{\Gamma(\frac{1}{2} -
i \omega)} } \sim \exp( i (\omega\log \omega - \omega) + {\cal
O}(1/\omega)) .
\ee
The stationary phase is at $\omega=\hat{\lambda}_{\rm in}
\hat{\lambda}_{\rm out}$ which is indeed where we expect the
energy to be located, and the expression for the energy becomes
\be
E = \frac{\partial}{\partial (-i \log (\hat{\lambda}_{\rm in}
\hat{\lambda}_{\rm out} ))} \log (\exp(-i \hat{\lambda}_{\rm in}
\hat{\lambda}_{\rm out})) = \hat{\lambda}_{\rm in}
\hat{\lambda}_{\rm out}.
\ee
This is indeed the correct energy for an orbit of the type
\be
 \lambda(x) \sim \hat{\lambda}_{\rm in} e^{-t} +
\hat{\lambda}_{\rm out} e^t .
\ee
We can easily take the fluctuation determinant into account in
this discussion. This merely adds an extra term $i/2$ to the energy,
from which we see that this is indeed a subleading effect,
since $\hat{\lambda}_{\rm in}$ and $\hat{\lambda}_{\rm out}$
both contain a factor of $\sqrt{\mu}$ that we absorbed at the beginning
of this calculation.

\subsection{Adding tachyons}

The construction of the tachyon operators uses crucially an
asymptotic bosonization procedure. To add tachyons to the game,
we need the normal ordered products
\be \rho(x,t) = :\psi^{\dagger}(x,t) \psi(x,t):
\ee
in the same in and out limits as above. This operator in boson
language will become $\partial \phi$ or $\bar{\partial} \phi$.
Thus we can read off the tachyon operators by looking at modes of
$\rho(x,t)$. The normal ordering is with respect to
the Fermi surface, but this is only relevant for the zero energy tachyon which
we will not consider.

It is straightforward to write down the tachyon operators from
the normal ordered product of two fermions. We have the operators
\bea
T^{\dagger}_{{\rm in},k} & = & \int d\omega \Psi_{{\rm in},\omega+k}
\Psi^{\dagger}_{{\rm in},\omega}
\nonumber \\
T_{{\rm out},k} & = & \int d\omega \Psi_{{\rm out},\omega-k}
\Psi^{\dagger}_{{\rm out},\omega} .
\eea
These have the property that for $k>0$, $T^{\dagger}_{{\rm in},k}$ creates a
positive energy incoming tachyon of momentum $k$, whereas $T_{{\rm out},k}$
acting on $\langle \mu |$ creates an out state with a positive energy outgoing
tachyon of momentum $k$. Clearly, $T^{\dagger}_{{\rm in},k}=T_{{\rm in},-k}$ and
similarly for $T_{{\rm out}}$.
The commutation relations are indeed those of a free scalar
\be
[T_{{\rm in},k} , T^{\dagger}_{{\rm in},k'} ] =
[T_{{\rm out},k} , T^{\dagger}_{{\rm out},k'} ] = k \delta(k-k') .
\ee
Up to a rescaling, these are of course the same as the tachyon modes
that we wrote in (\ref{aux11}).

We now have defined brane creation and annihilation operators, and
tachyon creation and annihilation operators. We can now write down
a general amplitude involving branes (with prescribed open string tachyon
profiles) and closed string tachyons, and the matrix model result for such
an amplitude can simply be worked out used the commutation relations of
the $\Psi_{\rm in,out}$ operators, and the fact that
\bea
\Psi^{\dagger}_{{\rm in},\omega} |\mu\rangle
 =
\Psi^{\dagger}_{{\rm out},\omega} |\mu\rangle = 0 & & {\rm for}\,\,\,\,
\omega > \mu \nonumber \\
\Psi_{{\rm in},\omega} |\mu\rangle
 =
\Psi_{{\rm out},\omega} |\mu\rangle = 0 & & {\rm for}\,\,\,\,
\omega < \mu \nonumber .
\eea

Notice that it matters whether we first create a brane and then a tachyon, or
whether we first create a tachyon and then a brane,
\be
[D^{\dagger}_{{\rm in},\lambda}, T^{\dagger}_{{\rm in},k}] =
e^{-ik \log {\hat \lambda}_{{\rm in}}} D^{\dagger}_{{\rm in},\lambda}.
\ee
This is a signal of the fact that there is duality between bosons and fermions
(bosonization), or equivalently, that there is a duality between open and closed
strings. The complete Hilbert space can be spanned either by closed strings states
plus their solitons (the closed string = boson picture),
or by systems of branes and anti-branes (the open string = fermion picture).

To illustrate this picture, we will now compute a few quantities and
check the results against results in the literature and our cylinder
calculation described in section~(\ref{cylinder}).

Tachyon correlators have been computed in e.g. \cite{mooreplramg}
with precisely these conventions. The world-sheet tachyons are obtained by
including a leg-pole factor, e.g.
\be
T^{\dagger}_{{\rm in},\omega; {\rm world-sheet}} =
\frac{\Gamma(i\omega)}{\Gamma(-i\omega)} T^{\dagger}_{{\rm in},\omega} .
\ee
In the conventions of \cite{polchinski} we should include an extra factor
of $\sqrt{2\pi}$ in the right hand side of this expression. One also has the
option of including a factor $1/\mu$ on the right hand side, this is a
simple choice of normalization, and one could also include phases in the
definition but we will not do that here.

The first quantity we will consider is the emission probability of a tachyon
by a rolling tachyon. For this purpose we need to compute
\be
\frac{ \left\langle \mu | D_{{\rm out},\hat{\lambda}_{\rm out}}
T_{{\rm out},k} D^{\dagger}_{{\rm in},\hat{\lambda}_{\rm in}}
| \mu \right\rangle}{
 \left\langle \mu | D_{{\rm out},\hat{\lambda}_{\rm out}}
D^{\dagger}_{{\rm in},\hat{\lambda}_{\rm in}}
| \mu \right\rangle} .
\ee
This can easily be computed using Wick contractions, and we find
\be
-\frac{
\int_{-\infty}^{{\rm min}(\mu,\mu+k)} d\omega e^{-i\omega\log\hat{\lambda}_{\rm out}
- i (\omega-k) \log\hat{\lambda}_{\rm in}} R_{\omega-k} }{
\int_{-\infty}^{\mu} d\omega e^{-i\omega\log\hat{\lambda}_{\rm out}
- i \omega \log\hat{\lambda}_{\rm in}} R_{\omega}
} .
\ee
If we compute numerator and denominator in the stationary phase approximation, which
is the leading contribution in the
$1/\mu$ expansion as discussed above, we find for the emission
probability
\be
{\cal A} = -e^{-i k \log\hat{\lambda}_{\rm out}} .
\ee
Up to a prefactor which depends on the choice of convention, this is
the answer one expects \cite{klebanovmaldacenaseiberg} for the emission
probability for a tachyon from the decaying brane.
Inserting $T^{\dagger}_{\rm out}$ would naturally describe the absorption
of the tachyon by a decaying brane.
We could also have chosen the order of the out brane and out tachyon operator
the other way around. Then the results of the calculation
would have no tree level contribution.

A similar calculation yields for the tree-level contribution to the
emission  amplitude
\be
\frac{ \left\langle \mu | D_{{\rm out},\hat{\lambda}_{\rm out}}
T_{{\rm in},k}^{\dagger} D^{\dagger}_{{\rm in},\hat{\lambda}_{\rm in}}
| \mu \right\rangle}{
 \left\langle \mu | D_{{\rm out},\hat{\lambda}_{\rm out}}
D^{\dagger}_{{\rm in},\hat{\lambda}_{\rm in}}
| \mu \right\rangle}  =  -e^{-i k \log\hat{\lambda}_{\rm in}}
\ee
Inserting $T_{{\rm in}}$ again would describe
absorption of a tachyon by an incoming brane.

\subsection{Two branes}

Finally,
we consider the situation with two D-branes. From the
boundary state point of view, we have one in-boundary state
corresponding to one D-brane, and another out-boundary state
corresponding to the complex conjugated D-brane. Therefore, we
propose that the relevant matrix model calculation that describes
the interaction between two D-branes is \be \label{aux22} {\cal A}
= \langle D_{{\rm out},\hat{\lambda}^1_{\rm out}} D_{{\rm
in},\hat{\lambda}^2_{\rm in}} D^{\dagger}_{{\rm
out},\hat{\lambda}^2_{\rm out}} D^{\dagger}_{{\rm
in},\hat{\lambda}^1_{\rm in}} \rangle
\ee
The operators $ D_{{\rm in},\hat{\lambda}^2_{\rm in}}$ and
$D^{\dagger}_{{\rm out},\hat{\lambda}^2_{\rm out}}$ that
describe the second brane have been complex conjugated, so
that the role of in and out has been interchanged, and their definition
therefore involves phases $e^{i\omega \log\hat{\lambda}^2_{\rm in}}$ and
$e^{i\omega \log\hat{\lambda}^2_{\rm out}}$.

The amplitude in question can be rewritten using Wick's theorem as
 \bea  \label{jj9} {\cal A}&=& \int_{-\infty}^{\mu} d\omega
R_{\omega}^{\ast} e^{ i \omega (\log\hat{\lambda}^2_{\rm in} +
\log\hat{\lambda}^2_{\rm out}) } \int_{-\infty}^{\mu} d\omega
R_{\omega} e^{-i \omega (\log\hat{\lambda}^1_{\rm in} +
\log\hat{\lambda}^1_{\rm out}) } \nn &-& \int_{-\infty}^{\mu}
d\omega  e^{ i \omega (\log\hat{\lambda}^2_{\rm in} -
\log\hat{\lambda}^1_{\rm in}) } \int_{-\infty}^{\mu} d\omega  e^{
i \omega (\log\hat{\lambda}^2_{\rm out} - \log\hat{\lambda}^1_{\rm
out}) }
 \eea
The first term in the expression has the interpretation as
interactions between the incoming and outgoing branes, the second
term gives the interaction between the incoming-incoming and
outgoing-outgoing branes. More precisely, in terms of the boundary
state computation, each integral corresponds to the interaction
between the relevant incoming/outgoing states.
The incoming/outgoing states are described by choosing
the relevant time contour, or picking the part of the full brane
computation. Thus the above four-point calculation gives the
``square'' of the cylinder. Also notice that the above result
contains full information of the interaction beyond cylinder
diagram. The information is contained in the reflection
coefficient $R_\omega$.

\begin{figure}[ht]
\begin{center}\mbox{\epsfysize=4cm\epsfbox{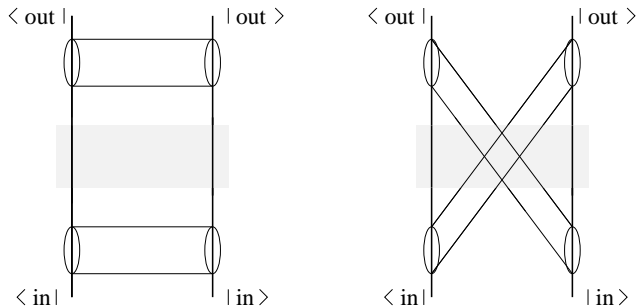}}\end{center}
\caption{\small Interactions of two D0-branes are composed of two terms
each of which contain two cylinders. The shaded area denotes the strongly
coupled region.}
\label{twobrane}
\end{figure}

To compare with the string theory calculation therefore we take a single
integral piece. Focusing on the piece in the second term which describes
the interaction of two incoming branes, we have
\be
\int_{-\infty}^{\mu} d\omega  e^{ i \omega
(\log\hat{\lambda}^2_{\rm in} - \log\hat{\lambda}^1_{\rm in}) }
\ee
which can be evaluated to be
\be
e^{ i \mu
(\log\hat{\lambda}^2_{\rm in} - \log\hat{\lambda}^1_{\rm in}) }
\left(
\pi \delta (\log\hat{\lambda}^2_{\rm in} - \log\hat{\lambda}^1_{\rm in}) -
\frac{i}{ \log\hat{\lambda}^2_{\rm in} - \log\hat{\lambda}^1_{\rm in} }
\right).
\ee
The presence of the $\d$-function is understood from the exclusion principle,
one cannot place two fermions on the same location. Assuming ${\hat \l_1}
$ is not equal $\hat{\l_2}$, and neglecting overlap with the Fermi-sea
we get for the cylinder diagram between two incoming branes
\be
\label{jj10}
\exp{\left( - \gamma_{{\rm E}} + \int_{0}^{\infty} \frac{dk}{k} e^{ik(
 \log\hat{\lambda}^2_{\rm in} - \log\hat{\lambda}^1_{\rm in}  )}
\right)} .
\footnote{Note that the integration is divergent when $k$
approaches zero. We regulate the divergence, and Euler's constant
$\gamma_{\rm E}$
appears by the process of regularization.} 
\ee
This result has the interpretation as the exponential
of a single string theory annulus diagram, in agreement with the
computations in section \ref{MatrixI}.
Note that the
full computation  of (\ref{jj9}) has the other piece
multiplied to (\ref{jj10}),
which has the interpretation as the annulus diagram between outgoing
states. Thus (\ref{jj9}) can be thought of as the combination of two
annulus diagrams as depicted in Figure~\ref{twobrane}.
The two point function calculated in section \ref{MatrixI} corresponds to
the contribution of one of the two cylinders. We see that the cylinders
are originated only from the asymptotic infinities. This is because
these cylinders are the genus suppressed ones as we explain in
Appendix A. Closed string modes emitted or absorbed at the strongly coupled
region will be affected by higher genus corrections.

Notice that (\ref{jj9}) is the exact answer in the
matrix model, and that again
we see no sign of divergences and/or the exchange of discrete states.

\section{Annulus diagram from boundary states} \label{cylinder}

\subsection{Preliminaries}

In this section, we consider the interaction of two unstable
D0-branes in terms of boundary state formalism, and compare 
with the results of the previous sections. Each brane carries
an open string tachyonic degree of freedom. The rolling of the
tachyon can be described by boundary conformal field theory with a
boundary interaction
\be \label{halfs}
S = {\l \over 2} \int d t e^{X^0 (t)}
\ee
This
interaction describes a decaying (half) brane solution, where the
tachyon rolls down its potential. One can also consider a bounce
(full brane) solution with a $\cosh$ boundary interaction
\be \label{fulls} S =
\l \int d t \cosh{X^{0}(t)}
\ee
This perturbation can be interpreted as a brane building up from a
finely tuned incoming coherent tachyon state, and decaying again
by rolling down the potential.

In the matrix model, the decaying solution is interpreted as an
eigenvalue rolling down the potential, while the bouncing solution
as an eigenvalue coming in from asymptotic infinity, rolling up
the potential, then rolling down again. The precise interpretation
of these boundary states in terms of free fermions was discussed
in detail in section \ref{MatrixII}.

The boundary state in the two dimensional string theory decomposes
into an $X^{0}$ dependent (rolling tachyon) part, a Liouville
part, and a ghost part \be |B \rangle = |B \rangle_{X^{0}} \otimes
|B \rangle_{L} \otimes |B \rangle_{{\rm ghost}} \ee The unstable
D0-branes are described by the ZZ boundary states in the Liouville
theory \cite{FZZ,ZZ}. The rolling part of the boundary state is
less well understood, and we first turn to a brief discussion of
the rolling part of the boundary state.

\subsection{The boundary state for the rolling tachyon}

We are interested in determining the boundary state that describes
the boundary interaction (\ref{halfs}) in the free theory that
consists of a timelike scalar. This is somewhat subtle issue. Two
alternative representations of this boundary state have been
proposed in the literature. In the first one we expand the
boundary state in Ishibashi states $|\o \rangle \!\rangle$ 
built out of normalizable
operators in the free timelike scalar theory. This representation
is naturally obtained by considering a spacelike Liouville theory
with arbitrary background charge, and by taking a limit where the
background charge vanishes. This leads to the boundary state
\cite{stromgut}
\be \label{rep1}
|B\rangle_{X^0}  = \int d\omega e^{-i \o \log\hat{\lambda}}
\frac{-i\pi}{\sinh \pi\omega} |\omega\rangle \!\rangle ,
\ee
where
\be
\hat{\lambda} = \pi\lambda .
\ee
There is an alternative representation of the boundary state,
obtained by Wick rotating the boundary state for an Euclidean
boson with boundary interaction $e^{iX}$; this boundary state is
an expansion in Ishibashi states built on non-normalizable
operators and its form can be obtained from \cite{callan,gabreck}
\be \label{rep2}
|B\rangle_{X^0}  = \sum_j \sum_{m\geq 0}^j \left(\begin{array}{c}
j+m \\ 2m \end{array} \right)  (i\hat{\lambda})^{2m} |j,m,m
\rangle \!\rangle .
\ee
Here the sum is over nonnegative half-integer $j$, and the sum
over $m$ is such that $m+j$ is an integer. For some more
discussion of these boundary states see e.g. \cite{okuyama} and
references therein. This representation of the boundary state was
used in the original work of Sen \cite{senroll}.

Both representations (\ref{rep1}) and (\ref{rep2}) are obtained by
indirect procedures, either taking a limit of Liouville theory or
by a suitable analytic continuation. One can also try to directly
compute the boundary state by considering disk diagrams with an
operator inserted in the middle of the dis, see e.g.
\cite{cola,stromgut,naqvi}. These calculations are in principle
well-defined, except for the contribution of the zero-mode, which
is where all the subtleties arise. If we represent the operator
inserted in the middle of the Dis through the state-operator
correspondence as
\be
V = e^{i\omega X^0}{\cal P}(\partial X^0,\partial^2
X^0,\ldots,\bar{\partial} X^0,\bar{\partial}^2 X^0,\ldots)
\ee
for some polynomial ${\cal P}$, then the disk calculation will
lead to an answer of the form (where $t$ is the zero mode of
$X^0$)
\be \label{intrep}
\int dt e^{i\omega t} \sum_{n=0}^{\infty} \hat{\lambda}^n e^{nt}
c_n({\cal P})
\ee
where $c_n({\cal P})$ does not depend on $t$ or $\hat{\lambda}$.
It represents the disk calculation with $n$ insertions of the
boundary operator and a zero momentum operator ${\cal P}(\partial
X^0,\bar{\partial} X^0,
\ldots)$ inserted in the middle of the disk. We can now see
how we have to treat (\ref{intrep}) in order to recover either
(\ref{rep1}) or (\ref{rep2}).

To recover (\ref{rep2}), we should apply momentum conservation to
(\ref{intrep}), and interpret the $t$ integral as
\be \int e^{i\omega t} e^{nt} dt \sim \delta(\omega - i n)
\ee
which one would naturally get by Wick rotating the zero mode and
viewing the integral as a Fourier integral. With this
interpretation, there is only a contribution for imaginary momenta
$\omega=i n$ and this leads to the boundary state (\ref{rep2}).
The coefficients $c_n({\cal P})$ have been computed for various
${\cal P}$ in \cite{cola}, leading for example to the results
\be \label{simex}
c_n(1) = (-1)^n, \qquad c_n(\partial X^0,\bar{\partial} X^0)
\sim (-1)^n - 2 \delta_{n,0}.
\ee
The $\delta_{n,0}$ that appears is directly related to the
nontrivial combination of Ishibashi states present in
(\ref{rep2}).

To recover (\ref{rep1}), we first consider ${\cal P}=1$. Then
(\ref{intrep}) becomes
\be \label{p1}
\int dt e^{i\omega t} \rho(\lambda,t)
\ee
with
\be \label{r1}
\rho(\lambda,t) = {1 \over 1 + \hat{\l} e^{t}}
\ee
In order to make sense of (\ref{p1}), we have to choose an
integration prescription. In \cite{liulambertmaldacena} the
integral is defined by  taking the contour over the real axis with
a small rotation $t \rightarrow t(1 - i \epsilon)$, to include the
poles of $\rho(t)$. With this definition we find
\be \label{p4}
\int dt e^{i\omega t} \rho(\lambda,t) = e^{-i\omega
\log\hat{\lambda}} \frac{-i\pi}{\sinh \pi\omega}
\ee
which is precisely what we have in (\ref{rep1}). However, we only
considered the case ${\cal P}=1$, and we should extend this
calculation to include all ${\cal P}$. One can see from simple
examples that this is not at all straightforward. For example,
taking ${\cal P}=\partial^2 X^0 \bar{\partial}^2 X^0$ \cite{cola}
leads in (\ref{intrep}) to
\be \label{p2}
\int dt e^{i\omega t} (\rho(\lambda,t)-2+\hat{\lambda} e^t)
\ee
and the contour integral given above does not yet give a good
definition of the integral $\int dt e^{i\omega t}  e^t$. One way
to define these integrals is to demand that the resulting boundary
state that we obtain is conformally invariant, i.e. obeys
\be \label{p5}
(L_n-\bar{L}_{-n}) |B\rangle_{X^0} =0.
\ee
Since the highest weight representations of the Virasoro algebra
with $c=1$ and conformal weight $-\omega^2<0$ do not contain any
null vectors\footnote{The Kac determinant for $c=1$ theories at
level $N$ is $\Delta_N(h,c)\sim \prod_{pq\leq N} (h
-(p-q)^2/4)^{P(N-pq)}$ and clearly this has no zeroes for $h<0$.
}\,\footnote{The absence of null-vectors is related to the statement
in \cite{liulambertmaldacena} that one can choose a gauge at
nonzero energy where the timelike oscillators are not present.},
it is sufficient to find the disk diagrams with ${\cal P}=1$ only;
all other disk diagrams then follow from (\ref{p5}). Thus,
conformal invariance of the boundary state plus the prescription
(\ref{p4}) provide a unique answer for all ill-defined integrals
such as (\ref{p2}), and the resulting answer is then given by
(\ref{rep1}). All this is true except at zero momentum. For
$\omega=0$, the corresponding highest weight representation has
null vectors and (\ref{p5}) is not sufficient to determine the
boundary state. This will not play an important role in what
follows, but we note that the contribution to (\ref{rep2}) with
zero momentum is the sum of the terms with $m=0$, so it simply is
$\sum_j |j,0,0\rangle \!\rangle$, which behaves as if there were no
null vectors anyway.

We have now motivated the two representations (\ref{rep1}) and
(\ref{rep2}) from direct world-sheet calculations. It appears that
some of the information of (\ref{rep2}) got lost in relating it to
(\ref{rep1}); in particular, we could have taken the zero mode
approximation of (\ref{rep2}) as in \cite{senroll}, and from that
we could also have derived (\ref{rep1}) via a Fourier
transformation. On the other hand, the zero mode structure of
(\ref{rep2}) plus conformal invariance determine (\ref{rep1}) up
to the contribution of $\omega=0$, so perhaps it should be
possible to somehow re-obtain (\ref{rep2}) from (\ref{rep1}) as
well.

In this paper we are mainly interested in computing an annulus
diagram involving two boundary states. Though (\ref{rep1}) and
(\ref{rep2}) are related, it is not clear whether and how
calculations involving (\ref{rep1}) are identical to calculations
involving (\ref{rep2}). If we use (\ref{rep2}) one might be
inclined to write down the following expression for the annulus
amplitude
\be \label{q1}
{\cal A} \sim \sum_{j} \sum_{m\geq 0}^j (\hat{\lambda}_1
\hat{\lambda}_2)^{2m} \left( \begin{array}{c} j+m \\ 2m
\end{array} \right)^2 \frac{q^{j^2}-q^{(j+1)^2}}{\eta(q)} .
\ee
It is not clear that this is correct, since we assumed here that
the Ishibashi states $|j,m,m\rangle\rangle$ are orthonormal to
each other. The zero-mode contribution to the overlap of two such
Ishibashi states involves integrals of the form $\int dt
e^{(n-n')t}$ which is not well-defined. Taking it equal to
$\delta_{n,n'}$ leads to (\ref{q1}). However, it is not obvious
that this the correct thing to do. Another peculiar feature of
(\ref{q1}) is that it contains binomial coefficients that grow
very rapidly. All these problems are related to the fact that
(\ref{rep2}) is an expansion in non-normalizable states, and a
general framework to formulate string amplitudes in terms of such
states does not yet exist.

The boundary state (\ref{rep1}) on the other hand is a much more
conventional expression in terms of normalizable Ishibashi states.
It is this type of expansion that we would normally use when
computing string theory amplitudes, and we will use (\ref{rep1})
in the rest of this section. It turns out that (\ref{rep1}) yields
answers that can be directly compared to matrix model
calculations. It remains to be seen whether and how the same
results can be recovered from (\ref{rep2}).

Except for the half-brane (\ref{halfs}) we will also be interested
in the full-brane (\ref{fulls}). For the full brane a discussion
similar to the one above applies, the main difference being that
$\rho(\lambda,t)$ in (\ref{r1}) gets replaced by
\be
\rho(\l, t) = {1 \over 1 + \hat{\l} e^{t}} + {1 \over 1 -
  \hat{\l} e^{-t}} - 1, \quad \hat{\l} = \sin{\l \pi} .
\ee
For the full brane we need to choose in (\ref{p4}) an integration
contour. One can choose a real contour, which corresponds to a
brane building up from a finely tuned tachyon state, and then
decaying back. Another contour choice is the Hartle-Hawking
contour, consisting of a piece along the imaginary time axis from
$t \in (0, i \infty)$, and a real part $t \in (0, \infty)$ (or $t
\in (-\infty, 0))$. These correspond to a preparing a state at
$t=0$ and evolving it to (from) the asymptotic infinity
\cite{liulambertmaldacena}.

\subsection{Cylinder computation}

We now concentrate on computing the cylinder diagram of two decaying
D0-branes. In terms of boundary states it can be written as
\be
A_{{\rm ZZ}} = {1 \over 2} \int_{0}^{1} {d q \over q}
\bzzl \otimes \langle B_{\l'}|
 q^{2 L_0} \blr \otimes \bzzr {\rm A}_{{\rm ghost}}
\ee
In this expression $\blr$ is the time dependent part of the boundary state,
and $\bzzr$ is the Liouville part. The ghost contribution is separated into
$A_{{\rm ghost}}$. We consider the overlap of two different rolling
boundary states, parametrized by $\l$ and $\l'$. Also, $q$ is the modulus
of the annulus. Since the interacting states are time-dependent, we compute the
full time-integrated diagram.

The Liouville part of the annulus diagram is already given in
\cite{ZZ,martinec},
\be
Z_{1,1} = \int d p \chi_{p} \Psi_{1,1} (p) \Psi_{1,1} (-p)
\ee
where the index $\{1,1\}$ refers to the fact that
we take the $m=1$, $n=1$ boundary state,
$\Psi_{1,1}$ is the wave function, and the character is
\bea
\chi_p &=& {q^{p^2} \over \eta(\tau)} \nn
\eta(\tau) &=& q^{{1\over 24}} \prod_{n=1}^{\infty} (1 -q^n)
\eea
More explicitly
\be
Z_{1,1} = -\int {d p \over 2 \pi i} {q^{p^2} \over \eta(q)}
\left({2 \over \pi}\sinh (\pi p)\right)^2
\ee
The ghost part of the annulus is
\be
A_{{\rm ghost}} = (\eta(q))^2
\ee
We still need to compute the time-dependent part. Inserting
Ishibashi states built over $e^{i \o X^{0}} |0\rangle$
\bea
A_{\l, \l'} &=& \langle B_{\l'}| q^{2 L_0} \blr =\int d t
\int d t' \int d \omega \int
d \o' \langle B_{\l'} |\o\rangle \langle \! 
\langle \o | q^{2 L_0} |\o'\rangle \! \rangle
\langle \o' \blr \nn \\
&=& \int d \omega {q^{-\o^2} \over \eta(q)} \rho(\l, t) \rho (\l', t')
e^{i \o (t - t')}
\eea

As we discussed in the previous subsection, we used for our
boundary state one that is expanded purely in terms of
normalizable Ishibashi states, and which is therefore determined
by the zero-mode structure of the corresponding Euclidean boundary
state, together with conformal invariance.

Assembling the full annulus amplitude and performing the integration
over the modulus $q$ we obtain
\be
A_{{\rm ZZ}} = -{1 \over 2}\int dt \int dt' \int { d p \over 2 \pi i}
 \int d \o \rho(\l, t) \rho (\l', t')
{(\sinh (\pi p))^2 \over p^2 - \o^2} e^{i \o (t - t')}
\ee
Note that the ghost contribution canceled the $\eta$-functions.
We now have a choice to proceed with the $\omega$ or with the $t$
integration first.
Let us proceed with the $\omega$ integration,
\be
A_{{\rm ZZ}} = -
\int dt \int dt' \int {d p \over  p} \rho(\l, t) \rho (\l', t')
\left({\sinh (\pi p)\over \pi}\right)^2  e^{i p (t -t')}
\ee

Finally for the time integrations the contour has to be specified. The
integral of $\rho(t)$ has poles at discrete values on the imaginary axis.
For the half-brane we follow the prescription described in the previous
section, with the result
\be
\label{ft}
\int dt \rho(t) e^{i p t} = {-i \pi \over \sinh{\pi p}}
e^{-i p \log{\hat{\l}}}
\ee
So finally we arrive at
\be
A_{{\rm ZZ}} = \int {d p \over  p} e^{-i p (\log{\hat{\l}} - \log{\hat{\l'}})}
\label{to}
\ee
This result is in agreement with the matrix model computations 
\ref{LL} and \ref{jj10}. 

If we consider full-brane or bounce solution with real contour,
the annulus amplitude is modified to
\bea
 A_{{\rm ZZ}} = \int {d p \over  p}
 \left(e^{-i p \log{\hat{\l}}}-e^{+i p \log{\hat{\l}}}\right)
\left(e^{+i p \log{\hat{\l'}}}-e^{-i p \log{\hat{\l'}}} \right)
\eea

Had we done the $t$ integration first, we would have obtained instead
\be
A_{{\rm ZZ}} = {\pi^2 \over 2} \int d \o {d p \over 2 \pi i}
{1 \over (p^2 - \o^2)} \left( {\sinh \pi p \over \sinh{\pi \o}} \right)^2
e^{-i \o (\log{\hat{\l}} - \log{\hat{\l'}})}
\ee
Here we have taken the half-brane case for simplicity.
Since the $p$ integral diverges, we now have to proceed with the $\o$
integration. The expression has double poles at discrete imaginary values of
$\o$, which indicates the presence of discrete modes.
Performing the $\o$ integration we get the previous result (\ref{to})
plus an infinite sum containing the contribution from the discrete modes.
This infinite sum with contributions from discrete modes
does not seem to be reflected in
the matrix model result. The expression (\ref{to}) contains only an exchange
of intermediate tachyons, and this result is in agreement with the matrix
model computations.

\section{Discussion} \label{discussion}

We have seen that the matrix model computations performed in the
free fermionic picture agree with the string theory cylinder
diagram in which only the tachyon exchange is taken into account.
The matrix model correlators in the tree level ($g_s \rightarrow 0$)
approximation do not indicate the presence of discrete modes.

On the other hand, one expects a sign of the presence of discrete
states, at least in Euclidean signature, where they correspond to
normalizable string states. In that case they are part of the full
spectrum, and carry important physical information such as
gravitational interactions. The discrete modes certainly can be
seen to be implicitly present in the matrix model through the
ground ring generators and the $w_{\infty}$ algebra \cite{witten}.

In the cylinder computation between the two D0-branes we have
encountered several difficulties. A better understanding of these
difficulties might provide interesting insights into the nature of
the matrix model, in particular regarding the role of the discrete
modes and its non-perturbative degrees of freedom. In particular,
we would like to know whether the cylinder calculation can be
reformulated purely in terms of (\ref{rep2}) or not.
Another confusion arises relating a Euclidean worldsheet 
to a Minkowski spacetime computation.

Another inherent difficulty with the cylinder computation is the
divergence of the $p$-integration. We have chosen to take integrals
in a particular order to avoid the divergences. This corresponds
to regulating the divergent integral in a specific way.
We have chosen the regularization to be in agreement with the matrix
model computation, but again, this regularization procedure
cuts off an infinite sum of discrete modes.
Clearly, a better understanding of the string theory computation is
desirable.

There are several ways in which the results in this paper can
be extended. First of all, we can consider the interactions of
$p>2$ branes. From the matrix model point of view, this
can also be understood as taking $(N+p)\times(N+p)$ matrices,
integrating out the $N\times N$ degrees of freedom and taking a large
$N$ limit. According to \cite{DMP}, the result is given by a Kontsevich
matrix integral whose result is a certain $\tau$-function. This $\tau$-function
has an alternative representation as a determinant of fermion bilinears,
which is precisely what one obtains from a straightforward generalization
of (\ref{aux22}). This provides further evidence for the structure
advocated in section~\ref{MatrixII}. It would be interesting to understand
more directly how such a structure can emerge from world-sheet
calculations.

It is also relatively straightforward to generalize our results
to rolling tachyons in the $\hat{c}=1$ model of \cite{newhat}.
This simply amounts to filling both sides of the potential,
and incorporating both asymptotic regions of the matrix model.
In particular, we can study rolling tachyons corresponding to
perturbations of the form $\lambda \sinh X^0$, which start
at early times at $x\rightarrow -\infty$ and at late times
end up at $x\rightarrow +\infty$. It is easy to write down
corresponding expressions in the matrix model, but they are
more difficult to understand on the world-sheet, as these are
configurations that penetrate deeply in the non-perturbative
regime; they start out as coherent states of the symmetric
combination of the closed string tachyon and RR scalar, and end
up as coherent states of the antisymmetric combination instead.

Our calculations also clearly support Sen's point of view \cite{senopenclosed}
that open and closed strings are dual pictures of one and the same thing.
The matrix model as expressed in terms of the original variables (eigenvalues)
is the open string picture.
The fermionic description, which is a reformulation of the original matrix 
model, represents another formulation. The dual closed string picture arises 
from bosonization of the fermions.
In the closed string picture there is a clear distinction
between closed string states and D-branes: the second are coherent
states of the first (in other words, they are solitons in the closed
string). Similarly, in the open string picture closed strings arise
as excitations of brane/antibrane systems with total brane number zero.
Thus, both in the closed as well as in the open picture, there is
a conserved fermion$=$brane number, which is conserved despite the
fact that the branes are unstable.

\subsection*{Acknowledgement}
We would like to thank
Robbert Dijkgraaf, Vladimir Kazakov, Asad Naqvi, Soo-Jong Rey
and Seiji Terashima for helpful discussions. This work is
supported in part by the Stichting FOM.

\section*{Appendix A. Stationary phase limit of the two point function}

In this appendix, we consider the full two point correlator of the
free fermions without taking any asymptotic bosonization.
At the genus zero limit, we can take the stationary phase approximation.
This approximation extracts the piece of the correlator
corresponding to the tree level ($g_s \rightarrow 0$) string
computation discussed in section (\ref{cylinder}).

The full fermionic two point correlator is \cite{moore}
\bea
\langle \mu| T(\psi(t_1, x_1) \psi(t_2, x_2) |\mu\rangle^{E} &=&
\int d \o \left( \th(\o - \mu) \th(\Delta t)
- \th(\mu - \o) \th(-\Delta t) \right) e^{-\o \Delta t} \nn
&\times&(\psi^{+}(\o, x_1) \psi^{+}(\o, x_2) +
\psi^{-}(\o, x_1) \psi^{-}(\o, x_2))
\eea
The subscript $E$ refers to taking the Euclidean two point function.
Using the properties of wave functions of the inverse harmonic oscillator, or
by analytically continuing the harmonic oscillator propagator we arrive
at the equivalent expression \cite{moore}
\bea
\langle \mu| T(\psi(t_1, x_1) \psi(t_2, x_2) |\mu\rangle^{E}
&=& i e^{- \mu \Delta t} \int {d p \over 2 \pi} e^{-i p \Delta t}
\int_{0}^{{\rm sgn(p)}\cdot \infty}
d s \frac{e^{- s p  + i \mu s}}{(-4 \pi i \sinh{s})^{1/2})}
\nn
&\times& \exp{\left( -\frac{i\left((x_1^2 + x_2^2) \cosh{s}
- 2 x_1 x_2 \right)}{4 \sinh{s}}\right)}
\eea

To consider tree level interaction, we suppress the higher genus
corrections.
We rescale $x \rightarrow {\beta}^{1/2} x$, and
$\mu \rightarrow \beta \mu$, where $\b$ is sent
to infinity and $\beta \mu$ is kept fixed but large.
Taking this limit suppresses both loop corrections
(these are effectively $\hbar= 1/\b$ corrections), and tree
level ($1/{\b \mu}$) higher order interactions.
The first corrections would correspond
to the higher genus corrections in the string diagrams. The tree level
corrections would come from taking into account the interactions
of the bosonized scalar field. Since in this paper we are mostly interested
in the cylinder computation on the string side, we keep $\beta \mu \gg 1$.

Rescaling $x$ the integral acquires a large phase, and can be evaluated
with the stationary phase method. The phase will be stationary at
\be
e^{{\bar s}} + e^{-{\bar s}} = {x_1 \over x_2} + {x_2 \over x_1}
\ee
where $\bar{s}$ denotes the stationary value of $s$.
Depending on the signature of $p$ in the integration and on the signature
of $(x_1 - x_2)$, this gives
\bea
{\bar s}&=&\Th(p) \left(\Th(x_1 - x_2) \log{({x_1 \over x_2})} +
\Th(x_2 - x_1)\log{({x_2 \over x_1})} \right) \nn
&+& \Th(-p) \left(\Th(x_1 - x_2) \log{({x_2 \over x_1})} +
\Th(x_2 - x_1)  \log{({x_1 \over x_2})} \right)
\eea
Here $\Th$ is the step function.
At the stationary point part of the Gaussian factor coming
from the integration of the quadratic part cancels the
$1/\sqrt{\sinh{s}}$ factor in the integral. The rest of the expression
assembles to the WKB wave function, and an additional exponential factor,
giving
\bea
S(t_1, t_2, x_1, x_2)^E &=&
i \sqrt{2 \over \b |x_1 x_2|}e^{- \b \mu \Delta t}
\int_{0}^{\infty}{d p \over 2 \pi} e^{-i p \Delta t} \left\{\Th(p) \left[
\Th(x_1 -x_2) \times \right. \right. \nn
&&\left. \left. e^{-\frac{i \b}{4} (x_1^2 - x_2^2)}
e^{(i \b \mu - p) \log{x_1 \over x_2}} +
\Th(x_2 -x_1) e^{-\frac{i \b}{4} (x_2^2 -x_1^2)}
e^{(i \b \mu - p) \log{x_2 \over x_1}} \right] \right.\nn
&+&\Th(-p) \left[
\Th(x_1 -x_2) e^{-\frac{i \b}{4} (x_2^2 -x_1^2)}
e^{(i \b \mu - p) \log{x_2 \over x_1}} \right. \nn
&+& \left. \left. \Th(x_2 -x_1)
e^{-\frac{i \b}{4} (x_1^2 - x_2^2)}
e^{(i \b \mu - p) \log{x_1 \over x_2}}
\right] \right\}
\eea
Going back to Minkowski signature requires $\Delta t \rightarrow i \Delta t$,
$p \rightarrow i p$. The four pieces of the integral correspond to picking
the left or right moving pieces in the correlator:
$\Psi_{L, R}^{\dagger} \Psi_{L, R}$.
Dropping the WKB factor and choosing the left-left piece we
obtain for the two point correlator
\be
S(t_1, t_2, x_1, x_2) = \int_{0}^{\infty} d p e^{-i p \left(\log{x_1 \over x_2}
- \Delta t \right)}
\ee

Taking the orbits $x_{1,2} = {\hat \l}_{1,2} \cosh{t}$  (where t is sent to
infinity) we obtain the result (\ref{LL}) computed from the asymptotic
bosonization.

Notice that to get this result, we have only suppressed
higher genus and higher order tree level correction, we did not
consciously remove any
possible discrete modes.

\section*{Appendix B. Two point function of eigenvalue density}

In this appendix we consider two point function of the eigenvalue density.
Roughly speaking, since for chiral fermions
$\psi(x,t) \sim \exp\left(\int \psi^\dagger \psi (x,t)\right)$,
by computing the two point function of the eigenvalue density we can
try to see once more whether discrete modes are exchanged between two
D-branes, and we can also present an alternative derivation of the
cylinder diagram

The two point function of eigenvalue density is defined in terms of
\bea
\label{deftwo}
 G(t_1,x_1;t_2,x_2) &=& \langle\mu|\psi^\dagger\psi(t_1,x_1)\psi^\dagger
 \psi(t_2,x_2)|\mu\rangle \nn
 G(\omega_1,x_1;\omega_2,x_2) &=& \int dt_1 dt_2 e^{i\omega_1 t_1}
  e^{i\omega_2 t_2} G(t_1,x_1;t_2,x_2).
\eea
We can compute this using the two point functions of fermions computed
in appendix~A,
\bea
\label{twopoint}
 G(\omega_1,x_1;\omega_2,x_2) =&& i^2 \delta(\omega_1+\omega_2)
 \int_0^\infty ds \int_{-\infty}^\infty dt {1 \over s+t}
 e^{-s \omega_1} e^{i \mu (s+t)}  \nn
 &&
 \langle x_1|e^{2isH}|x_2\rangle \langle x_2|e^{2itH}|x_1\rangle,
\eea
where $\langle x_m|e^{2isH}|x_n\rangle$ defined as
\bea
 \langle x_m|e^{2isH}|x_n\rangle=\frac{1}{(-4 \pi i \sinh{s})^{1/2})}
 \exp{\left( -\frac{i\left((x_m^2 + x_n^2) \cosh{s}
  - 2 x_m x_n \right)}{4 \sinh{s}}\right)}
\eea
Here we have taken the Euclidean time. As in the Appendix A,
We will think of the genus zero limit, scaling $\mu \rightarrow \beta \mu,
x \rightarrow \sqrt{\beta} x$ and taking the limit $\beta \rightarrow
\infty$ with $ \beta \mu = {\rm fixed} \gg 1$.
Then we can apply the stationary phase approximation to the $s$ and $t$
integration.
But as before, there are two sets of saddle points :
$\bar s=\bar t=\pm \log (x_1 / x_2)$.
To have a non-vanishing denominator in (\ref{twopoint}), we choose the
same sign for the saddle of $\bar s$ and $\bar t$. Finally, we get
\bea
 G(\omega_1,x_1;\omega_2,x_2) = \delta(\omega_1+\omega_2)
 {1 \over 2 x_1 x_2 \log (x_2 / x_1)}
  e^{(2 i \mu - \o_1) \log(x_2 / x_1)
 -{i \over 2}(x_2^2-x_1^2)} .
\eea
We take the inverse Fourier transformation to $(t,x)$ coordinate space
\bea
 G(t_1,x_1;t_2,x_2) = \int d\omega_1 d\omega_2 e^{-i\omega_1 t_1}
 e^{-i \omega_2 t_2} G(\omega_1,x_1;\omega_2,x_2)
\eea

Stripping off WKB wavefunctions, and
Wick rotating to Minkowski space,  we get
\bea
 \label{midstep}
  \int d\omega {1 \over \log x_1-\log x_2}
  e^{-i \omega (t_1-\log x_1 - t_2 + \log x_2)}.
\eea

To get the cylinder diagram between two D-branes, we should integrate
over $t$ in (\ref{midstep}) up to a large nonzero value of
$t$, and then insert the classical
orbit at the boundary value of $t$, which yields the simple answer
\bea
 {\cal A} = \int_0^\infty
  {d\omega \over \omega} e^{-i \omega \log({\hat \lambda}_1
  /{\hat \lambda}_2)}.
\eea
Since
$\psi(x_1,t_1)\psi(x_2,t_2)
\sim \exp\left(\int \psi^\dagger \psi (x_x,t_1) \right)
\exp\left(\int \psi^\dagger \psi (x_2,t_2)\right)$, by
summing the all the ladders, we recover (\ref{LL}).
Note that we again see no divergences due to discrete mode exchange
(or any other sign of discrete modes for that matter).

We can evaluate the two point function of the eigenvalue density in another
way. If we take the Fourier transformation
from
$x$ to $z$,  and rotate $z$ to
$il$, we get the two point function of
macroscopic loop operators of loop length $l$. The result is already
well-known in the literature \cite{ml,moore}.
At genus zero the result is
\bea
 \langle W(w_1, l_1) W (w_2, l_2) \rangle = \delta(\omega_1+ \omega_2)
   \int {d \zeta \over \pi^2}
   {\zeta^2 \over \zeta^2 + \omega_1^2} K_{i\zeta}( - 2 \sqrt{\mu} l_1)
   K_{i\zeta}( - 2 \sqrt{\mu} l_2)
\eea
We can now analytically continue back to $z = -il$ and do an inverse
Fourier transformation to get the following expression in terms of $x$,
\bea
 &&{\pi \over \sqrt{x_1^2- 4 \mu}  \sqrt{x_2^2- 4\mu} } \int d \omega_1 
\omega_1 e^{-i \omega_1 (t_1 - t_2)} \times \nn
 &&  
 \cosh \left(2 \omega_1 \sinh^{-1} { \sqrt{{x_1 \over \sqrt{\mu}}-2}
 \over 2}\right)
 \cosh \left(2 \omega_1 \sinh^{-1} { \sqrt{{x_2 \over \sqrt{\mu}}-2}
 \over 2}\right).
\eea
To get the Minkowski result we perform a
Wick rotation $t \rightarrow - i t$ and
$ \omega \rightarrow i\omega$.
Again taking off the WKB factors, we have
\bea
 \sim \int d\omega_1 \omega_1 e^{-i \omega_1 (t_1 -t_2 - \log x_1 + \log x_2)}.
\eea
Also here we should do the integral over $t$ to reproduce the correct answer
for the cylinder diagram,
\bea
 {\cal A} &=& \int_0^\infty dt_1 dt_2 \int d \omega_1 \omega_1 e^{-i \omega_1 (t_1 -t_2 - \log x_1 + \log x_2)} \nn
 &=& \int { d \omega \over \omega} e^{-i \omega \log ({\hat \lambda}_1 /{\hat \lambda}_2)}.
\eea
In the last line, we have again inserted the classical trajectory at asymptotic
infinity. We find that the result agrees completely with the previous
results.

\end{document}